\documentclass[12pt]{amsart}
\pdfoutput=1
\usepackage{graphicx}
\usepackage{caption}
\usepackage{subcaption}
\usepackage{amsmath}
\usepackage{amssymb}
\usepackage{amsfonts}
\usepackage{amscd}
\usepackage{verbatim}
\usepackage{enumerate}
\usepackage{multirow}

\begin{document}

\title{Pseudoalignment for metagenomic read assignment}
\author[L. Schaeffer]{Lorian Schaeffer}
\address{Department of Molecular and Cell Biology, UC Berkeley}
\author[H. Pimentel]{Harold Pimentel}
\address{Department of Computer Science, UC Berkeley}
\author[Nicolas Bray]{Nicolas Bray}
\address{Innovative Genomics Initiative, UC Berkeley}
\author[P. Melsted]{P\'{a}ll Melsted}
\address{University of Iceland and Decode Genetics, Iceland}
\author[L. Pachter]{Lior Pachter}
\address{Departments of Mathematics, Molecular and Cell Biology \& Computer Science, UC Berkeley; {\tt lpachter@math.berkeley.edu}}
\begin{abstract}
We explore connections between metagenomic read assignment and the
quantification of transcripts from RNA-Seq data. In particular, we show that the recent idea of pseudoalignment introduced in the RNA-Seq context is suitable in the metagenomics setting. When coupled with the Expectation-Maximization (EM) algorithm, reads can be assigned far more accurately and quickly than is currently possible with state of the art software.
\end{abstract}
\maketitle

\section{Introduction}
The analysis of microbial communities via whole-genome shotgun
sequencing has led to exceptional bioinformatics challenges \cite{chen2005bioinformatics} that remain largely unsolved \cite{scholz2012next}. Most of these challenges can be characterized as ``{\it de
  novo}'' bioinformatics problems: they involve assembly of sequences,
binning of reads, and annotation of genes directly from sequenced
reads. The emphasis on {\it de novo} methods a decade
ago was the result of a paucity of sequenced reference microbial and
archaeal genomes at the time. However this has begun to change in recent years \cite{land2015insights}. As sequencing costs have
plummeted, the number of fully sequenced genomes has increased dramatically,
and while a large swath of the microbial world remains
uncharacterized, there are now thousands of ``reference quality'' genomes suitable for the application of reference-based methods.

One of the fundamental metagenomics problems that is amenable to reference-based analysis is that of ``sequence classification'' or ``read assignment''. This is the problem of assigning sequenced reads to taxa. The MEGAN program \cite{huson2007megan} was one of the first reference-based read assignment programs and was published shortly after sequencing-by-synthesis methods started to become mainstream. It provided a phylogenetic context to mapped reads by assigning reads to the lowest taxonomic level at which they could be uniquely aligned, and became popular in part because of a powerful accompanying visualization toolkit. One of the drawbacks of MEGAN was that its approach to assigning ambigously mapping reads limited its application to quantification of individual strains, an issue which was addressed in a number of subsequent papers, culminating in GRAMMy \cite{xia_grammy:_2011} and GASiC \cite{lindner_gasic:_2013}, which were the first to statistically assign ambigously mapped reads to individual strains. Unfortunately, these approaches all relied on read alignment, a computational problem that is particularly difficult in the metagenomic setting where reference genome databases are large and read sets gigantic.

In a breakthrough publication in 2014 \cite{wood_kraken:_2014} it was shown that it is possible to greatly accelerate read assignment utilizing fast k-mer hashing to circumvent the need for read alignment. An implementation called Kraken was used to show that analyses that previously took hours were tractable in minutes, and the removal of the read alignment step greatly simplified workflows and storage requirements. However the Kraken speed came at a cost. An examination of the Kraken algorithm and output reveals that the method takes a step back from GRAMMy and GASiC by discarding statistical assignment of reads at the strain level in favor of direct taxonomic assignment as in MEGAN (notably according to \cite{lindgreen2015evaluation}, Kraken is more accurate than MEGAN although, as we'll see, it is not as accurate as GASiC). A natural question to ask is whether the strengths of Kraken and GASiC can be combined, i.e. whether  it is possible to leverage fast k-mer based hashing to map reads not at the taxonomic but at the strain level, while assigning the resulting ambigously mapped reads using a statistical framework that allows for probabilistic assignment of reads.

To answer this question we turned to RNA-Seq \cite{cloonan2008stem,lister2008highly,nagalakshmi2008transcriptional,mortazavi2008mapping}, an experiment for which there has been extensive methods development that we hypothesized could be adapted and applied to metagenomics. Many of the challenges of metagenomic quantification translate to problems in RNA-Seq via a dictionary that replaces genome targets with transcript targets. For example, ambigously mapped genomic reads that are difficult to resolve at the strain level in the metagenomics setting are analogous to reads that are difficult to assign to specific isoforms in RNA-Seq. Statistical questions at the heart of ``comparative metagenomics'' \cite{huson2009methods, rodriguez2006application, tringe2005comparative} are analogous to the statistical problems in differential expression analysis. In fact, the only significant differences between metagenomics and RNA-Seq are that genome sequences are much larger than transcripts and reference databases are less complete. These differences have engineering implications, but statistically and computationally, metagenomics and transcriptomics are very much the same.

In this paper we show that technology transfer from RNA-Seq to metagenomics makes it possible to perform read assignment both rapidly {\em and} accurately. Specifically, we show that it is possible to accurately assign reads at the {\em strain} level using a fast k-mer based approach that goes beyond the hashing of Kraken and takes advantage of the principle of pseudoalignment \cite{bray2015near}. The idea of pseudoalignment originates with RNA-Seq, where it was developed to take advantage of the fact that the sufficient statistics for RNA-Seq quantification are assignments of reads to transcripts rather than their alignments. The same applies in the metagenomics setting, and we show that just as in RNA-Seq, application of the EM algorithm to ``equivalence classes'' \cite{nicolae2011estimation} allows for accurate statistical resolution of mapping ambiguities. Using a published simulated dataset \cite{mende_assessment_2012} and an implementation of pseudoalignment coupled to the EM algorithm in kallisto \cite{bray2015near}, we demonstrate significant accuracy and performance improvements in comparison to state of the art programs.

\section{Results}

To test the hypothesis that RNA-Seq quantification methods can be applied in the metagenomics setting we began by examining the performance of eXpress, a program that implements a streaming EM algorithm for RNA-Seq read assignment from alignments, on simulated data \cite{roberts2013streaming}. We chose eXpress because it utilizes traditional read alignments directly to a transcriptome but is more memory efficient than other approaches (e.g. RSEM \cite{li2011rsem}) and therefore more suitable in the metagenomics setting. Other RNA-Seq quantification tools such as Cufflinks \cite{trapnell2010transcript} were not suitable for our needs because of their dependence on read alignments to genomes and not transcriptomes, a requirement that does not translate easily to the metagenomics setting.

To test eXpress we aligned a simulated dataset of Illumina-like reads from 100 microbial genomes to a reference database containing only those genomes, allowing us to compare results to a ground truth (the Illumina100 data) \cite{mende_assessment_2012}. We began by comparing eXpress to GASiC, which also utilizes read alignments for read assignment. The results are shown in Table 1. We found that eXpress outperforms GASiC at the exact genome, species, genus, and phylum levels, which we believe is because the statistical model of eXpress takes into account data-dependent read error profiles in assigning reads.

A major problem with GASiC and eXpress is that the alignments they require are slow to generate. The alignments, made with Bowtie2 \cite{langmead2012fast}, took days. As reported in \cite{wood_kraken:_2014}, Kraken was much faster on the data, taking only 22 minutes 38s. We also tested CLARK \cite{ounit_clark:_2015}, another recently published k-mer based assignment tool and, in agreement with the benchmarks in \cite{lindgreen2015evaluation}, we found it to be slightly faster taking 20 minutes 30s. However, as seen in Table 1, both Kraken and CLARK have significantly worse performance than both GASiC and eXpress (in concordance with \cite{lindgreen2015evaluation} but in contradiction to \cite{ounit_clark:_2015}, we found that Kraken is more accurate than CLARK).

We next turned to a comparison of kallisto with Kraken and CLARK using the Illumina100 simulation (i100) but utilizing a larger and more realistic reference database of 1,958 genomes from \cite{martin_optimizing_2012}. The results, shown in Tables 1,2 and in Figure 1 (where the database is called ``i100+Martin'') show that kallisto is much more accurate than Kraken and CLARK at all four taxonomic levels tested. The performance of kallisto at the exact genome level is significantly better than Kraken at the genus level. Notably, Kraken's performance at the exact genome level (AVGRE 18.67 and RRMSE 38.26) is too poor to be of practical use.

{\tiny
\begin{table}[!h]
\centering
\begin{tabular}{lrrrrrrrr}
 & \multicolumn{2}{c}{Exact Genome} & \multicolumn{2}{c}{Species} & \multicolumn{2}{c}{Genus} & \multicolumn{2}{c}{Phylum} \\
 & \multicolumn{1}{c}{AVGRE} & \multicolumn{1}{c}{RRMSE} & \multicolumn{1}{c}{AVGRE} & \multicolumn{1}{c}{RRMSE} & \multicolumn{1}{c}{AVGRE} & \multicolumn{1}{c}{RRMSE} & \multicolumn{1}{c}{AVGRE} & \multicolumn{1}{c}{RRMSE} \\
\textit{i100} &  &  &  &  &  &  &  &  \\
\textbf{kallisto} & \textbf{0.97} & \textbf{5.42} & \textbf{0.14} & \textbf{0.36} & \textbf{0.13} & \textbf{0.38} & \textbf{0.09} & \textbf{0.10} \\
Kraken & 18.67 & 38.26 & 8.06 & 21.88 & 5.27 & 16.66 & 3.33 & 4.84 \\
CLARK & \multicolumn{1}{c}{--} & \multicolumn{1}{c}{--} & 12.28 & 22.73 & 10.32 & 18.22 & 7.52 & 7.88 \\
GASiC & 7.21 & 19.31 & 3.80 & 10.46 & 3.72 & 11.43 & 2.52 & 3.10 \\
eXpress & 2.57 & 11.92 & 0.40 & 0.61 & 0.34 & 0.57 & 0.13 & 0.18 \\
\textit{i100+Martin} &  &  &  &  &  &  &  &  \\
\textbf{kallisto} & \textbf{3.78} & \textbf{12.50} & \textbf{0.41} & \textbf{0.76} & \textbf{0.33} & \textbf{0.67} & \textbf{0.24} & \textbf{0.25} \\
Kraken & 38.00 & 56.28 & 10.76 & 26.81 & 3.62 & 13.45 & 1.34 & 2.24 \\
CLARK & \multicolumn{1}{c}{--} & \multicolumn{1}{c}{--} & 22.75 & 29.72 & 20.20 & 24.46 & 12.88 & 14.16
\end{tabular}
\vskip 0.1in
\caption{Normalized count based classification accuracy at four taxonomic ranks. CLARK results are missing at the strain level because the program does not output strain level counts.}
\end{table}}

The running time of kallisto is faster than Kraken with the i100 dataset (5m55s vs. 22m38s using a single core) but slower on the i100+Martin (50m55s vs. 27m3s). However kallisto index building is slightly faster than Kraken, and when utilizing multiple threads for pseudoalignment, kallisto's running time is negligible.

{\tiny
\begin{table}[!h]
\centering
\begin{tabular}{lrrrrrrr}
 & \multicolumn{1}{c}{\multirow{2}{*}{\begin{tabular}[c]{@{}c@{}}\%\\ Unmapped\end{tabular}}} & \multicolumn{2}{c}{Species} & \multicolumn{2}{c}{Genus} & \multicolumn{2}{c}{Phylum} \\
 & \multicolumn{1}{c}{} & \multicolumn{1}{c}{Precision} & \multicolumn{1}{c}{Sensitivity} & \multicolumn{1}{c}{Precision} & \multicolumn{1}{c}{Sensitivity} & \multicolumn{1}{c}{Precision} & \multicolumn{1}{c}{Sensitivity} \\
\textbf{kallisto} & \textbf{2.04\%} & \textbf{0.997} & \textbf{0.997} & \textbf{0.998} & \textbf{0.998} & \textbf{0.999} & \textbf{0.999} \\
Kraken & 2.62\% & 0.986 & 0.950 & 0.985 & 0.967 & 0.995 & 0.994 \\
Clark & 17.01\% & 0.864 & 0.864 & 0.884 & 0.884 & 0.977 & 0.977
\end{tabular}
\vskip 0.1in
\caption{Precision and sensitivity (see Methods) at three taxonomic ranks.}
\end{table}
}

While the results of kallisto on the i100 and i100+Martin databases were convincing, we decided to examine its performance in the case of missing strains, a situation that is commonplace in metagenomic analyses. We examined two different scenarios: the effect on performance when a strain is missing but other similar strains are in the database, and the case when a strain distant from others is missing. For the former we removed {\it Bacillus cereus ATCC1 0987}, leaving a number of other Bacilii, and for the latter we removed {\it Listeria welshimeri serovar 6b str. SLCC5334}, leaving no other Listerias.

{\tiny
\begin{table}[!h]
\centering
\begin{tabular}{lrrrrrrrr}
 & \multicolumn{2}{c}{Exact Genome} & \multicolumn{2}{c}{Species} & \multicolumn{2}{c}{Genus} & \multicolumn{2}{c}{Phylum} \\
 & \multicolumn{1}{c}{AVGRE} & \multicolumn{1}{c}{RRMSE} & \multicolumn{1}{c}{AVGRE} & \multicolumn{1}{c}{RRMSE} & \multicolumn{1}{c}{AVGRE} & \multicolumn{1}{c}{RRMSE} & \multicolumn{1}{c}{AVGRE} & \multicolumn{1}{c}{RRMSE} \\
\textit{No Listeria} &  &  &  &  &  &  &  &  \\
\textbf{kallisto} & \textbf{2.54} & \textbf{11.41} & \textbf{1.94} & \textbf{10.88} & \textbf{2.29} & \textbf{12.43} & \textbf{0.85} & \textbf{0.90} \\
Kraken & 20.12 & 39.58 & 9.80 & 24.46 & 7.43 & 20.80 & 3.78 & 5.01 \\
\textit{No Bacillus} & \multicolumn{1}{l}{} & \multicolumn{1}{l}{} & \multicolumn{1}{l}{} & \multicolumn{1}{l}{} & \multicolumn{1}{l}{} & \multicolumn{1}{l}{} &  &  \\
\textbf{kallisto} & \textbf{2.95} & \textbf{11.73} & \textbf{1.21} & \textbf{3.13} & \textbf{0.74} & \textbf{0.92} & \textbf{0.72} & \textbf{0.76} \\
Kraken & 19.61 & 39.28 & 8.45 & 21.74 & 5.65 & 16.72 & 3.60 & 4.65
\end{tabular}
\vskip 0.1in
\caption{Normalized count based classification accuracy after removing single genomes from index.}
\end{table}
}

As expected, Table 3 shows that the performance of both kallisto and Kraken degrades with removal of strains from the reference database. However, what is interesting is that in the case of Bacillus, read assignment is still possible at the genus level as reflected in improved performance. In the case of Listeria, the absence of neighboring species means that the reads cannot be assigned. Remarkably, even with a strain completely missing from the reference database, the performance of kallisto is still higher than that of Kraken with all genomes present (compare to Table 1).

\section{Methods}

\subsection{Illumina100 dataset}
We tested kallisto and alternate programs on a set of simulated reads published in \cite{mende_assessment_2012}. The Illumina100 dataset consists of 53.33 million 75bp reads, simulated by the iMESSi metagenomic simulator using an Illumina error model. The reads were simulated from a set of 100 unique bacterial genomes. The set is of genomes from 85 different species and 63 different genuses, over a range of abundances from 0.86\% to 2.2\%.

Reads were trimmed with the program Trimmomatic (version 0.32) \cite{bolger2014trimmomatic} to a minimum length of 40bp, using its adaptive trimming algorithm MAXINFO with a target length of 40 and default strictness. 40 reads were dropped due to quality issues.

\subsection{Taxonomic identification}
We analyzed each program's output at four taxonomic ranks: phylum, genus, species, and ``exact genome'' level. The latter tests the abundance estimation of the actual Illumina100 genomes, which are a combination of strains and substrains and thus aren't taxonomically well defined. The other three ranks are as assigned by NCBI's Taxonomy Database, as of November, 2015.

\subsection{Count estimation accuracy calculation}
Using a simulated dataset with known abundances allowed us to benchmark programs by comparing program outputs with true values for each genome. While kallisto is able to output length-corrected individual genome abundances, most of the programs we compared with only output counts, so for consistency we analyzed the accuracy of assigned or estimated counts for each program. We normalized the estimated counts by the percent of assigned reads in order to be able to compare relative count estimates between programs. 

We primarily used the error measures AVGRE (Average Relative Error), which computes the mean of the difference between truth and estimate, and RRMSE (Relative Root Mean Square Error), which computes the root mean square average of the difference between truth and estimate, to judge the accuracy of our estimates. Formally, with $n$ true genomes/species/genera/phyla, true counts $\tau_i$ ($1 \leq i \leq n$) and estimated counts $t_i$ at the rank, and $A$ aligned reads out of $T$ total reads we computed
\begin{eqnarray*}
AVGRE  &= & \frac{1}{n}\sum_{i}^n \frac{|t_i\cdot \frac{T}{A} - \tau_i|}{\tau_i}\quad \mbox{and} \\
RRMSE  & = & \sqrt{\frac{1}{n}\sum_{i}^n \left( \frac{t_i \cdot \frac{T}{A} - \tau_i}{\tau_i}\right)^2}.
\end{eqnarray*}

In addition, for comparison of programs using the metrics in \cite{wood_kraken:_2014}, we calculated the assignment aggregate precision and sensitivity of kallisto, Kraken and CLARK. Instead of examining the results of assignments of individual reads to specific species or genus, we relaxed the benchmark to instead measure sensitivity and precision based on aggregate counts at taxonomic ranks. Mimicking the computations of \cite{wood_kraken:_2014}, we computed the aggregate sensitivity at a given rank R by calculating (\# of counts correctly assigned at rank R)/(\# of counts assigned at rank R). Aggregate precision was calculated as (\# of counts correctly assigned at or below rank R)/(\# of counts assigned at or below rank R + \# of reads incorrectly assigned above rank R). Counts at a rank were considered to be correctly assigned when they were less than or equal to the true counts. Our level of granularity in assessing assignment reflects the quantity of interest in reference-based metagenomics, namely the accuracy of the aggregate number of counts assigned to individual members of a rank instead of the correctness of each individual read assignment.

The scripts used to compile the results are available at\\  {\tt https://github.com/pachterlab/metakallisto}

\subsection{Reference Genome Database}
In addition to aligning the Illumina100 reads against their originating genomes, we tested the more realistic case of aligning against a large bacterial database. In order to have a consistent, reproducible test that won't change as new bacteria are sequenced, we used a frozen published database for all our i100+Martin tests \cite{martin_optimizing_2012}. The database contains 1,751 bacterial genomes and plasmids spread over 1,253 species, in addition to 131 Archaea genomes. The published database also contained 3,683 viral genomes and 326 lower eukaryote genomes, but these were discarded before indexing. For compatibility with Kraken and CLARK, we also discarded 6 genomes that were lacking sequence GI numbers, and one header that didn't contain any sequence information (GI 308222630). This database was then combined with the Illumina100 source genomes for a total of 1,958 individual genomes.

\section{Discussion}

The idea of translating RNA-Seq methodology to and from metagenomics was, to our knowledge, first proposed in \cite{paulson2013differential} where statistical methods for identifying differential abundances in microbial marker genes were developed. In that paper, there were comparisons between the proposed metagenomics method and RNA-Seq differential analysis methods implemented in DESeq \cite{anders2010differential} and edgeR \cite{robinson2010edger}. Notably, the central idea of the paper, the specific consideration of zero inflated distributions to account for undersampling, is also used in single cell expression analysis \cite{mcdavid2013data}.

Our results show that RNA-Seq methods for quantification are also applicable in the metagenomics setting, and our results with kallisto demonstrate that it is possible to accurately and rapidly quantify the abundance of individual {\em strains}. With a few exceptions, e.g. \cite{bradley2015rapid}, most metagenomic analyses have focused on higher taxonomy, a point highlighted in the recent benchmarking paper \cite{lindgreen2015evaluation} which compares predictions at the phylum level because ``[comparisons at that level are] less prone to differences''. The phylum level is  four levels removed from genus, let alone species or strain. Our results suggests that the door is now open to metagenome analyses at the highest possible resolution.

While our benchmarks are based on simulated data, our experiments are much more realistic than previous analyses. For example, the Kraken and CLARK papers report results on simulations with ten genomes, whereas we have simulated from 100 genomes and mapped against nearly 2,000. One of the difficulties we faced in our analyses was the technical issue of taxonomic naming and annotation in collating results. This seemingly trivial matter is complicated by the lack of attention paid to low taxonomic level analysis in previous methods. Hopefully our results will spur others to standardize and organize naming conventions and analysis scripts so that low taxonomic level analysis can become routine.

In addition to quantification, we believe there is a lot of potential for differential analysis tools developed for RNA-Seq  to be applied more systematically in the metagenomics setting. In that regard, one of the interesting features of  kallisto is the ability to bootstrap to assess uncertainty in assignment, and we have utilized this to develop a new method and tool for differential analysis that takes advantage of the feature. The method should immediately be applicable in comparative metagenomics studies.

\section{Acknowledgments}

H.P. was supported by an NSF graduate research fellowship. P.M. was partially supported by a
Fulbright fellowship. L.S and L.P. were partially supported by NIH R01 HG006129 and NIH R01 DK094699.

\bibliographystyle{apalike}
\bibliography{metagenomics_paper.bib}

\begin{thebibliography}{}

\bibitem[Anders and Huber, 2010]{anders2010differential}
Anders, S. and Huber, W. (2010).
\newblock Differential expression analysis for sequence count data.
\newblock {\em Genome biol}, 11(10):R106.

\bibitem[Bolger et~al., 2014]{bolger2014trimmomatic}
Bolger, A.~M., Lohse, M., and Usadel, B. (2014).
\newblock Trimmomatic: a flexible trimmer for illumina sequence data.
\newblock {\em Bioinformatics}, page btu170.

\bibitem[Bradley et~al., 2015]{bradley2015rapid}
Bradley, P., Gordon, N.~C., Walker, T.~M., Dunn, L., Heys, S., Huang, B.,
  Earle, S., Pankhurst, L.~J., Anson, L., de~Cesare, M., et~al. (2015).
\newblock Rapid antibiotic resistance predictions from genome sequence data for
  {S. aureus and M. tuberculosis}.
\newblock {\em bioRxiv}, page 018564.

\bibitem[Bray et~al., 2015]{bray2015near}
Bray, N., Pimentel, H., Melsted, P., and Pachter, L. (2015).
\newblock Near-optimal {RNA-Seq} quantification.
\newblock {\em arXiv preprint arXiv:1505.02710}.

\bibitem[Chen and Pachter, 2005]{chen2005bioinformatics}
Chen, K. and Pachter, L. (2005).
\newblock Bioinformatics for whole-genome shotgun sequencing of microbial
  communities.
\newblock {\em PLoS Comput Biol}, 1(2):106--112.

\bibitem[Cloonan et~al., 2008]{cloonan2008stem}
Cloonan, N., Forrest, A.~R., Kolle, G., Gardiner, B.~B., Faulkner, G.~J.,
  Brown, M.~K., Taylor, D.~F., Steptoe, A.~L., Wani, S., Bethel, G., et~al.
  (2008).
\newblock Stem cell transcriptome profiling via massive-scale {mRNA}
  sequencing.
\newblock {\em Nature methods}, 5(7):613--619.

\bibitem[Huson et~al., 2007]{huson2007megan}
Huson, D.~H., Auch, A.~F., Qi, J., and Schuster, S.~C. (2007).
\newblock {MEGAN} analysis of metagenomic data.
\newblock {\em Genome research}, 17(3):377--386.

\bibitem[Huson et~al., 2009]{huson2009methods}
Huson, D.~H., Richter, D.~C., Mitra, S., Auch, A.~F., and Schuster, S.~C.
  (2009).
\newblock Methods for comparative metagenomics.
\newblock {\em BMC bioinformatics}, 10(Suppl 1):S12.

\bibitem[Land et~al., 2015]{land2015insights}
Land, M., Hauser, L., Jun, S.-R., Nookaew, I., Leuze, M.~R., Ahn, T.-H.,
  Karpinets, T., Lund, O., Kora, G., Wassenaar, T., et~al. (2015).
\newblock Insights from 20 years of bacterial genome sequencing.
\newblock {\em Functional \& integrative genomics}, 15(2):141--161.

\bibitem[Langmead and Salzberg, 2012]{langmead2012fast}
Langmead, B. and Salzberg, S.~L. (2012).
\newblock Fast gapped-read alignment with {Bowtie} 2.
\newblock {\em Nature methods}, 9(4):357--359.

\bibitem[Li and Dewey, 2011]{li2011rsem}
Li, B. and Dewey, C.~N. (2011).
\newblock {RSEM}: accurate transcript quantification from {RNA-Seq} data with
  or without a reference genome.
\newblock {\em BMC bioinformatics}, 12(1):323.

\bibitem[Lindgreen et~al., 2015]{lindgreen2015evaluation}
Lindgreen, S., Adair, K.~L., and Gardner, P. (2015).
\newblock An evaluation of the accuracy and speed of metagenome analysis tools.
\newblock {\em bioRxiv}, page 017830.

\bibitem[Lindner and Renard, 2013]{lindner_gasic:_2013}
Lindner, M.~S. and Renard, B.~Y. (2013).
\newblock {GASiC}: {Metagenomic} abundance estimation and diagnostic testing on
  species level.
\newblock {\em Nucleic Acids Research}, 41(1):e10.

\bibitem[Lister et~al., 2008]{lister2008highly}
Lister, R., O'Malley, R.~C., Tonti-Filippini, J., Gregory, B.~D., Berry, C.~C.,
  Millar, A.~H., and Ecker, J.~R. (2008).
\newblock Highly integrated single-base resolution maps of the epigenome in
  arabidopsis.
\newblock {\em Cell}, 133(3):523--536.

\bibitem[Martin et~al., 2012]{martin_optimizing_2012}
Martin, J., Sykes, S., Young, S., Kota, K., Sanka, R., Sheth, N., Orvis, J.,
  Sodergren, E., Wang, Z., Weinstock, G.~M., and Mitreva, M. (2012).
\newblock Optimizing {Read} {Mapping} to {Reference} {Genomes} to {Determine}
  {Composition} and {Species} {Prevalence} in {Microbial} {Communities}.
\newblock {\em PLoS ONE}, 7(6):e36427.

\bibitem[McDavid et~al., 2013]{mcdavid2013data}
McDavid, A., Finak, G., Chattopadyay, P.~K., Dominguez, M., Lamoreaux, L., Ma,
  S.~S., Roederer, M., and Gottardo, R. (2013).
\newblock Data exploration, quality control and testing in single-cell
  {qPCR}-based gene expression experiments.
\newblock {\em Bioinformatics}, 29(4):461--467.

\bibitem[Mende et~al., 2012]{mende_assessment_2012}
Mende, D.~R., Waller, A.~S., Sunagawa, S., Järvelin, A.~I., Chan, M.~M.,
  Arumugam, M., Raes, J., and Bork, P. (2012).
\newblock Assessment of {Metagenomic} {Assembly} {Using} {Simulated} {Next}
  {Generation} {Sequencing} {Data}.
\newblock {\em PLoS ONE}, 7(2):e31386.

\bibitem[Mortazavi et~al., 2008]{mortazavi2008mapping}
Mortazavi, A., Williams, B.~A., McCue, K., Schaeffer, L., and Wold, B. (2008).
\newblock Mapping and quantifying mammalian transcriptomes by {RNA-Seq}.
\newblock {\em Nature methods}, 5(7):621--628.

\bibitem[Nagalakshmi et~al., 2008]{nagalakshmi2008transcriptional}
Nagalakshmi, U., Wang, Z., Waern, K., Shou, C., Raha, D., Gerstein, M., and
  Snyder, M. (2008).
\newblock The transcriptional landscape of the yeast genome defined by {RNA}
  sequencing.
\newblock {\em Science}, 320(5881):1344--1349.

\bibitem[Nicolae et~al., 2011]{nicolae2011estimation}
Nicolae, M., Mangul, S., Mandoiu, I.~I., and Zelikovsky, A. (2011).
\newblock Estimation of alternative splicing isoform frequencies from {RNA-Seq}
  data.
\newblock {\em Algorithms for Molecular Biology}, 6(1):9.

\bibitem[Ounit et~al., 2015]{ounit_clark:_2015}
Ounit, R., Wanamaker, S., Close, T.~J., and Lonardi, S. (2015).
\newblock {CLARK}: fast and accurate classification of metagenomic and genomic
  sequences using discriminative k-mers.
\newblock {\em BMC Genomics}, 16(1):236.

\bibitem[Paulson et~al., 2013]{paulson2013differential}
Paulson, J.~N., Stine, O.~C., Bravo, H.~C., and Pop, M. (2013).
\newblock Differential abundance analysis for microbial marker-gene surveys.
\newblock {\em Nature methods}, 10(12):1200--1202.

\bibitem[Roberts and Pachter, 2013]{roberts2013streaming}
Roberts, A. and Pachter, L. (2013).
\newblock Streaming fragment assignment for real-time analysis of sequencing
  experiments.
\newblock {\em Nature methods}, 10(1):71--73.

\bibitem[Robinson et~al., 2010]{robinson2010edger}
Robinson, M.~D., McCarthy, D.~J., and Smyth, G.~K. (2010).
\newblock {edgeR: a Bioconductor package for differential expression analysis
  of digital gene expression data}.
\newblock {\em Bioinformatics}, 26(1):139--140.

\bibitem[Rodriguez-Brito et~al., 2006]{rodriguez2006application}
Rodriguez-Brito, B., Rohwer, F., and Edwards, R.~A. (2006).
\newblock An application of statistics to comparative metagenomics.
\newblock {\em BMC bioinformatics}, 7(1):162.

\bibitem[Scholz et~al., 2012]{scholz2012next}
Scholz, M.~B., Lo, C.-C., and Chain, P.~S. (2012).
\newblock Next generation sequencing and bioinformatic bottlenecks: the current
  state of metagenomic data analysis.
\newblock {\em Current opinion in biotechnology}, 23(1):9--15.

\bibitem[Trapnell et~al., 2010]{trapnell2010transcript}
Trapnell, C., Williams, B.~A., Pertea, G., Mortazavi, A., Kwan, G., Van~Baren,
  M.~J., Salzberg, S.~L., Wold, B.~J., and Pachter, L. (2010).
\newblock Transcript assembly and quantification by {RNA-Seq} reveals
  unannotated transcripts and isoform switching during cell differentiation.
\newblock {\em Nature biotechnology}, 28(5):511--515.

\bibitem[Tringe et~al., 2005]{tringe2005comparative}
Tringe, S.~G., Von~Mering, C., Kobayashi, A., Salamov, A.~A., Chen, K., Chang,
  H.~W., Podar, M., Short, J.~M., Mathur, E.~J., Detter, J.~C., et~al. (2005).
\newblock Comparative metagenomics of microbial communities.
\newblock {\em Science}, 308(5721):554--557.

\bibitem[Wood and Salzberg, 2014]{wood_kraken:_2014}
Wood, D.~E. and Salzberg, S.~L. (2014).
\newblock Kraken: ultrafast metagenomic sequence classification using exact
  alignments.
\newblock {\em Genome Biology}, 15(3):R46.

\bibitem[Xia et~al., 2011]{xia_grammy:_2011}
Xia, L.~C., Cram, J.~A., Chen, T., Fuhrman, J.~A., and Sun, F. (2011).
\newblock {GRAMMy}: {Accurate} {Genome} {Relative} {Abundance} {Estimation}
  {Based} on {Shotgun} {Metagenomic} {Reads}.
\newblock {\em PLoS ONE}, 6(12).

\end{thebibliography}

\begin{figure}[!ht]
    \begin{subfigure}[b]{\textwidth}
\vskip -0.2in
\includegraphics[width=\textwidth]{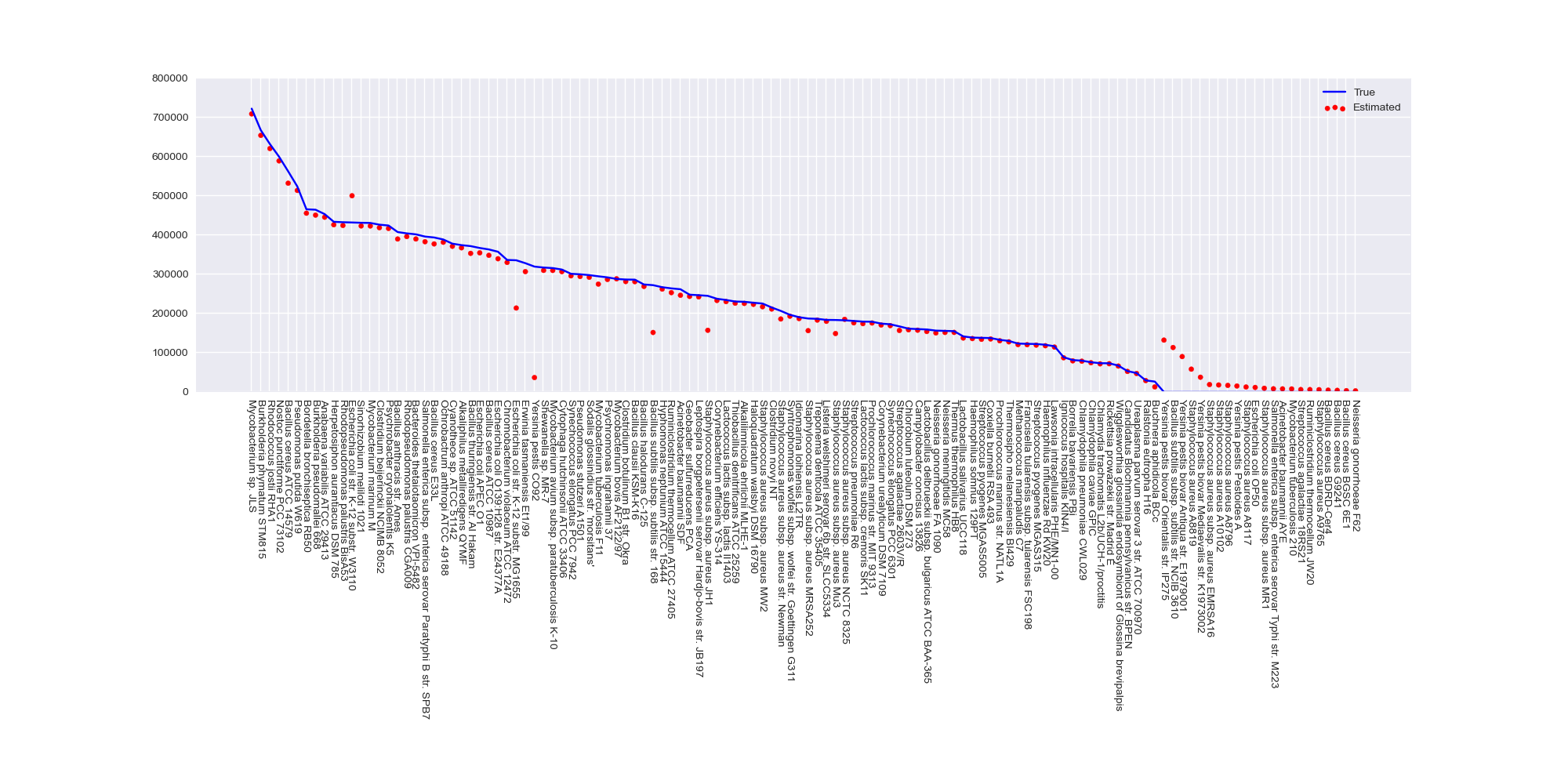}
       \label{fig:kallisto_strain}
\end{subfigure}
\vskip -0.4in
    \begin{subfigure}[b]{\textwidth}
\includegraphics[width=\textwidth]{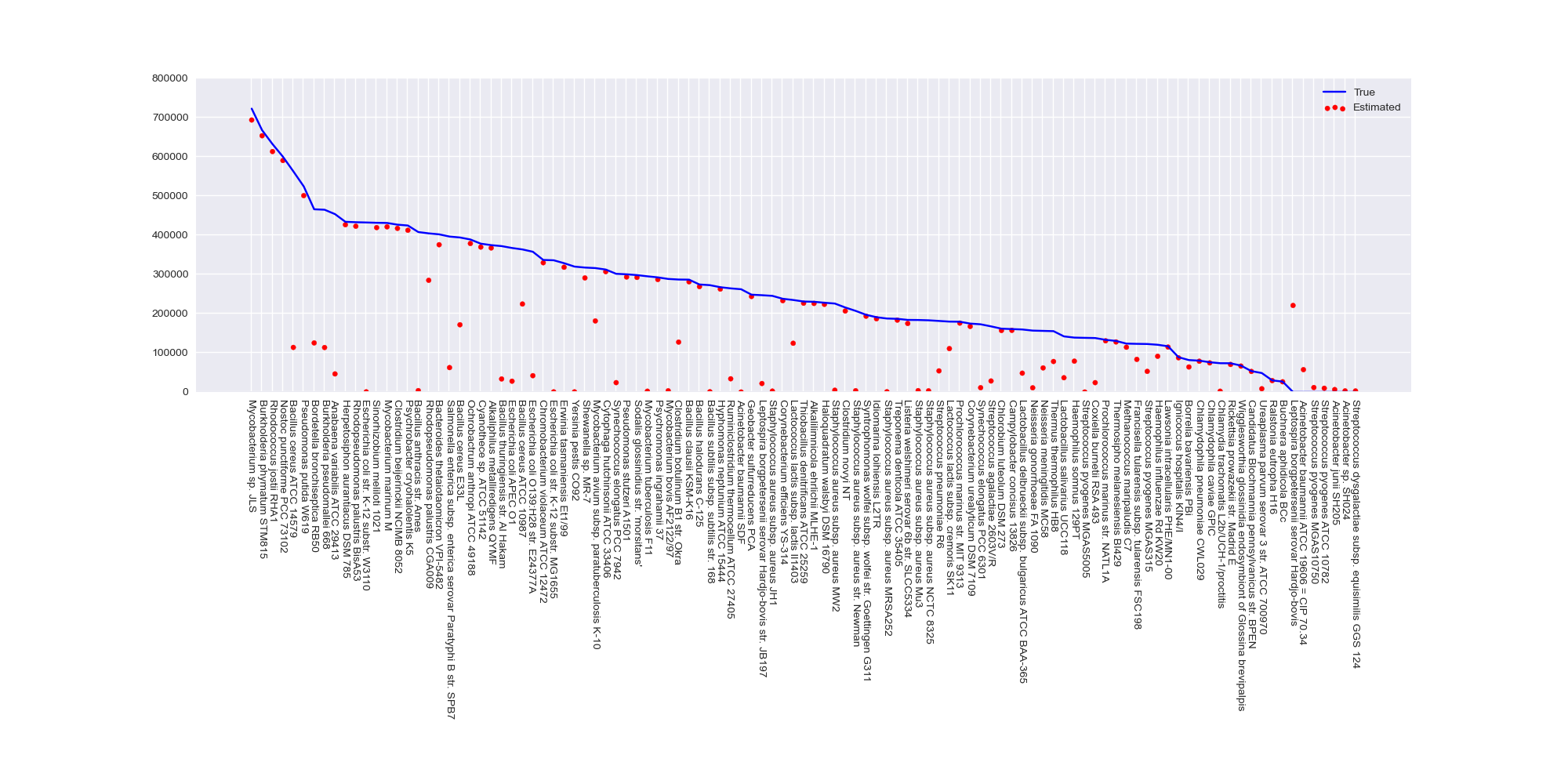}
       \label{fig:kraken_strain}
\end{subfigure}

\vskip -0.4in
\caption{Results of kallisto (top) and Kraken (bottom) on simulated reads from the i100+Martin dataset at the exact genome level.}
\vskip4in
\end{figure}

\begin{figure}[!ht]
    \begin{subfigure}[b]{\textwidth}
\vskip -0.2in
\includegraphics[width=\textwidth]{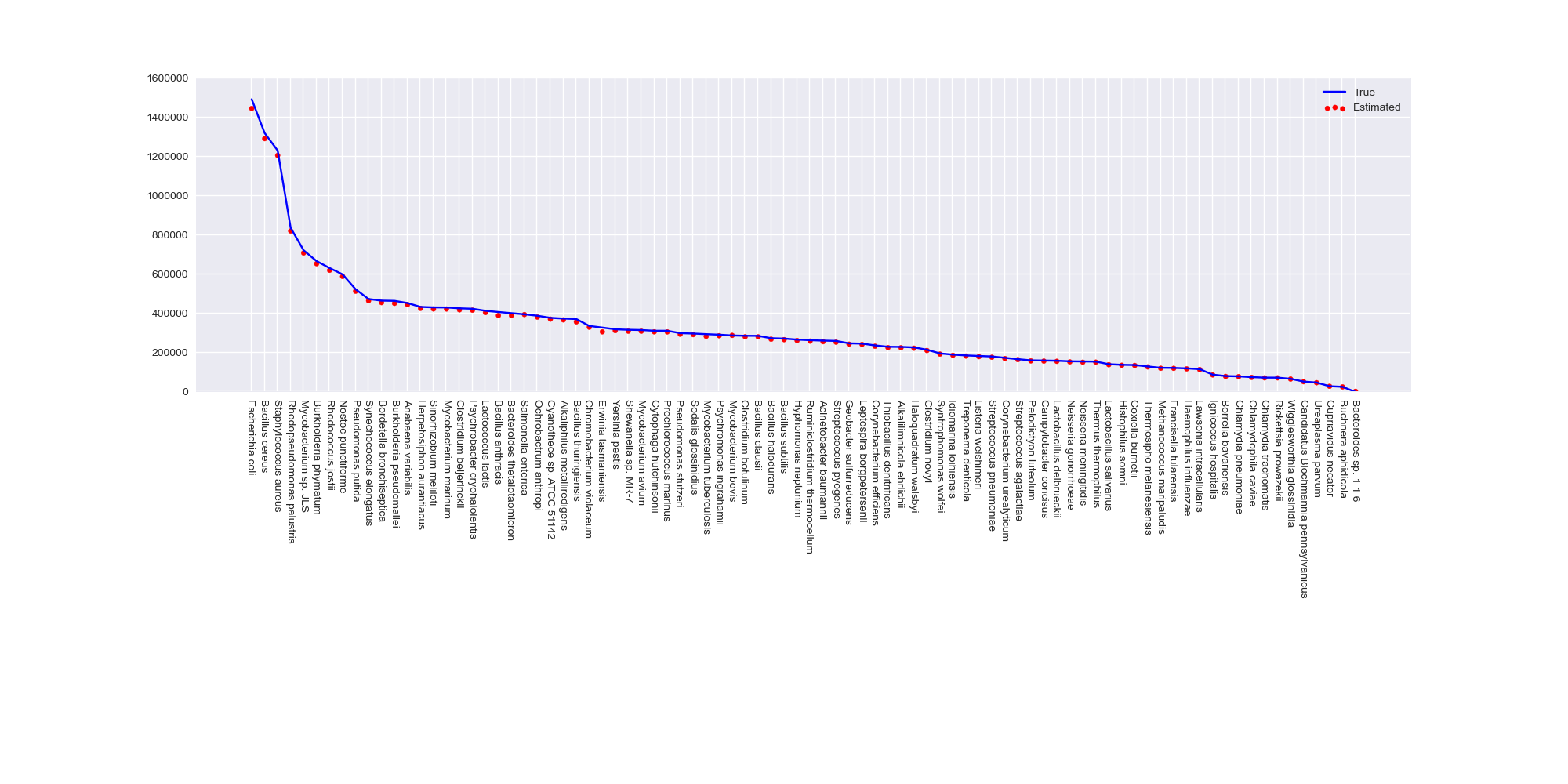}
       \label{fig:kallisto_species}
\end{subfigure}
\vskip -0.4in
    \begin{subfigure}[b]{\textwidth}
\includegraphics[width=\textwidth]{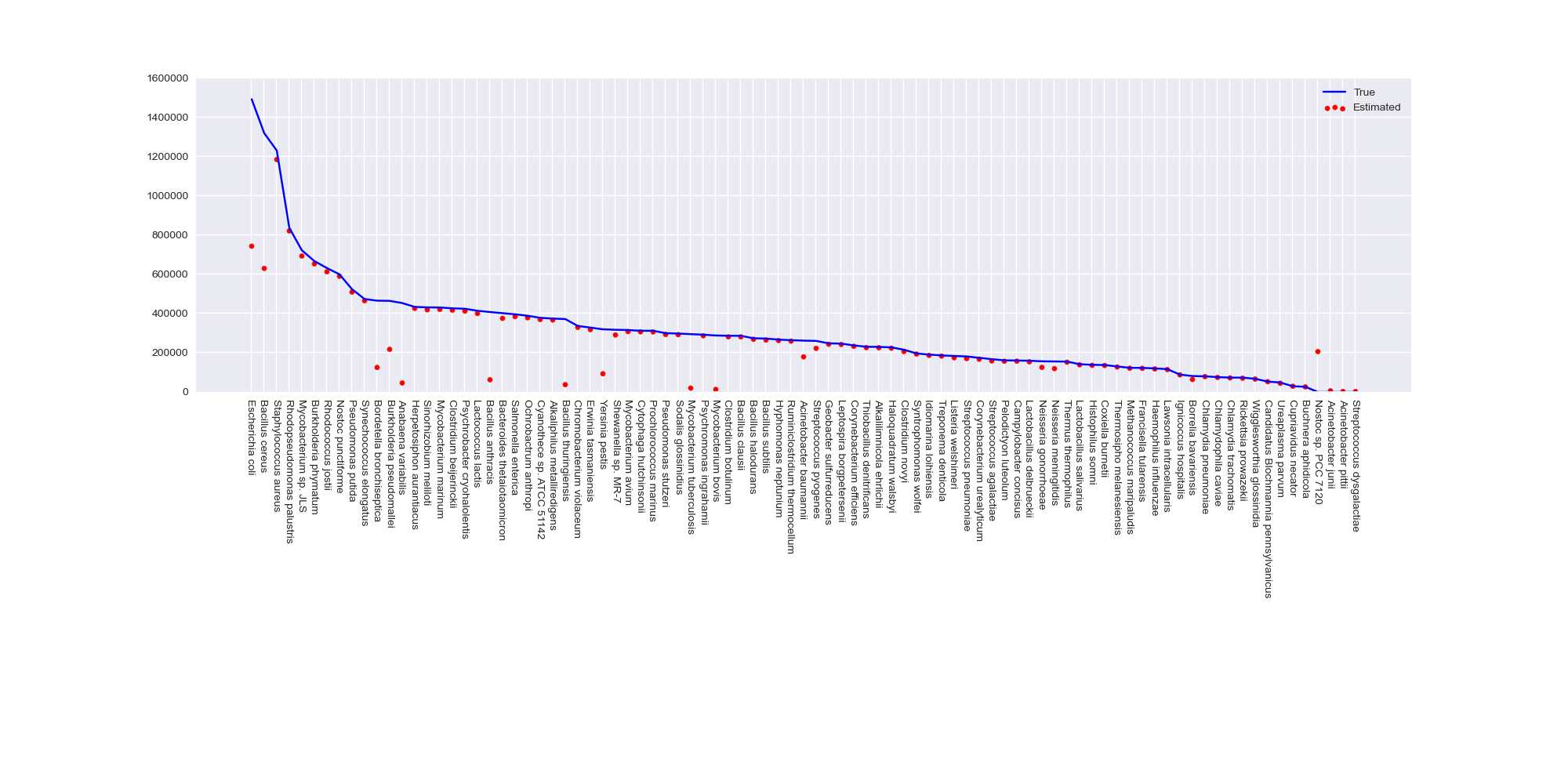}
       \label{fig:kraken_species}
\end{subfigure}
\vskip -0.4in
\begin{subfigure}[b]{\textwidth}
\includegraphics[width=\textwidth]{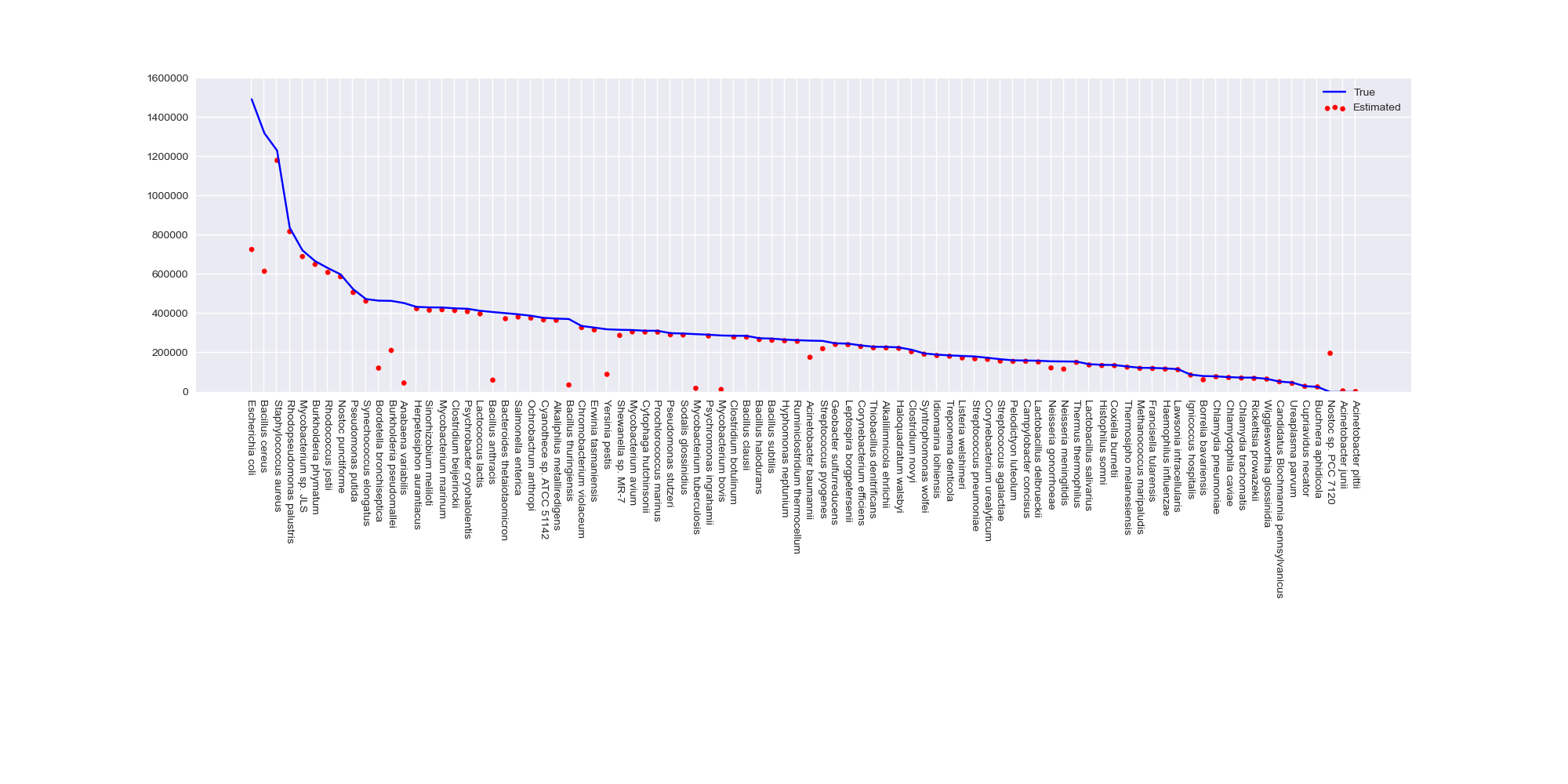}
       \label{fig:clark_species}
\end{subfigure}
\vskip -0.9in
\caption{Results of kallisto, Kraken and CLARK on simulated reads from the i100+Martin dataset at the species level.}
\end{figure}

\begin{figure}[!t]
    \begin{subfigure}[b]{\textwidth}
\vskip -0.2in
\includegraphics[width=\textwidth]{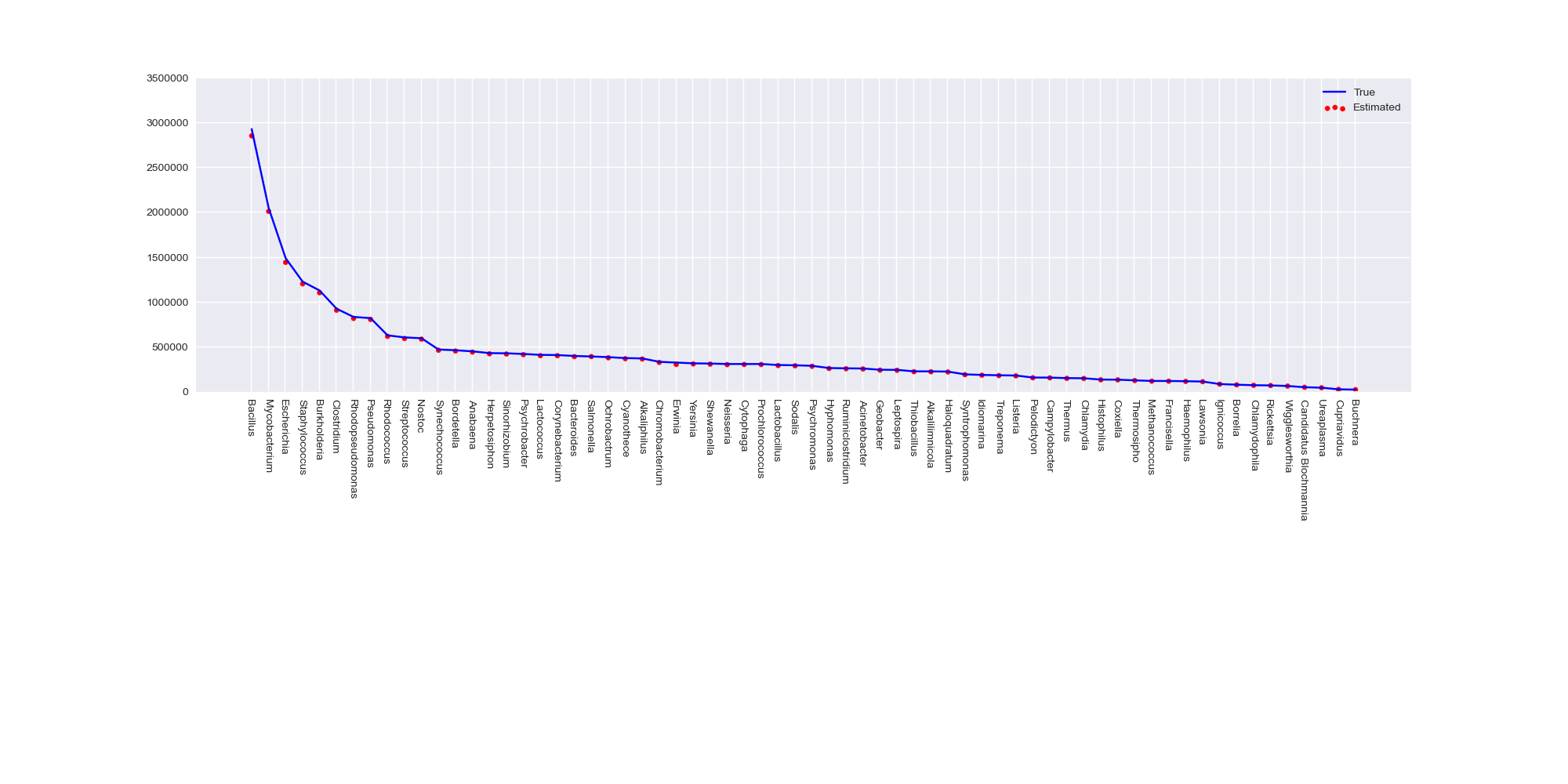}
       \label{fig:kallisto_genus}
\end{subfigure}
\vskip -0.4in
    \begin{subfigure}[b]{\textwidth}
\includegraphics[width=\textwidth]{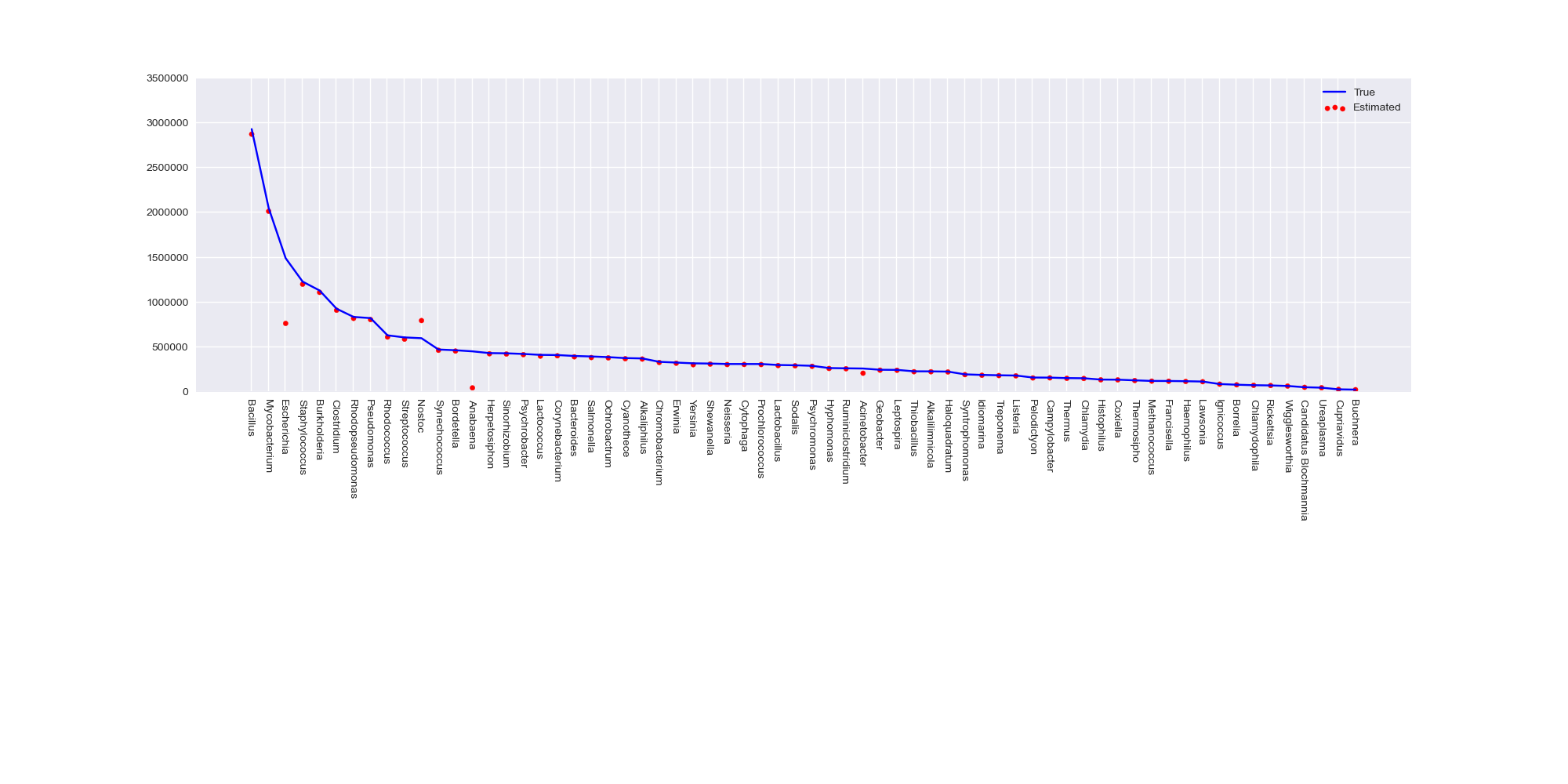}
       \label{fig:kraken_genus}
\end{subfigure}
\vskip -0.4in
\begin{subfigure}[b]{\textwidth}
\includegraphics[width=\textwidth]{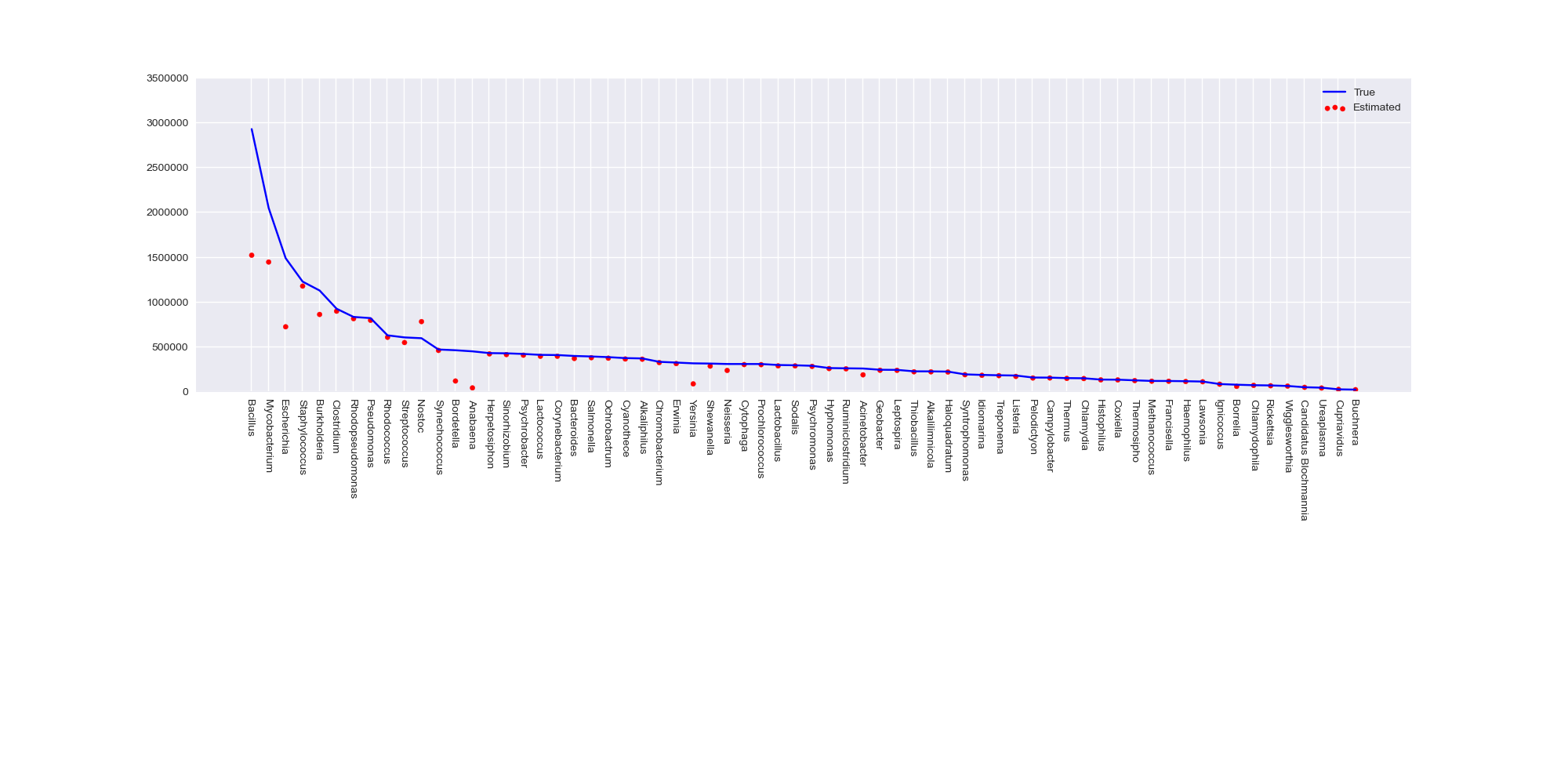}
       \label{fig:clark_genus}
\end{subfigure}
\vskip -1in
\caption{Results of kallisto (top), Kraken (middle) and CLARK (bottom) on simulated reads from the i100+Martin dataset at the genus level.}
\end{figure}

\begin{figure}[!t]
    \begin{subfigure}[b]{\textwidth}
\vskip -0.2in
\includegraphics[width=\textwidth]{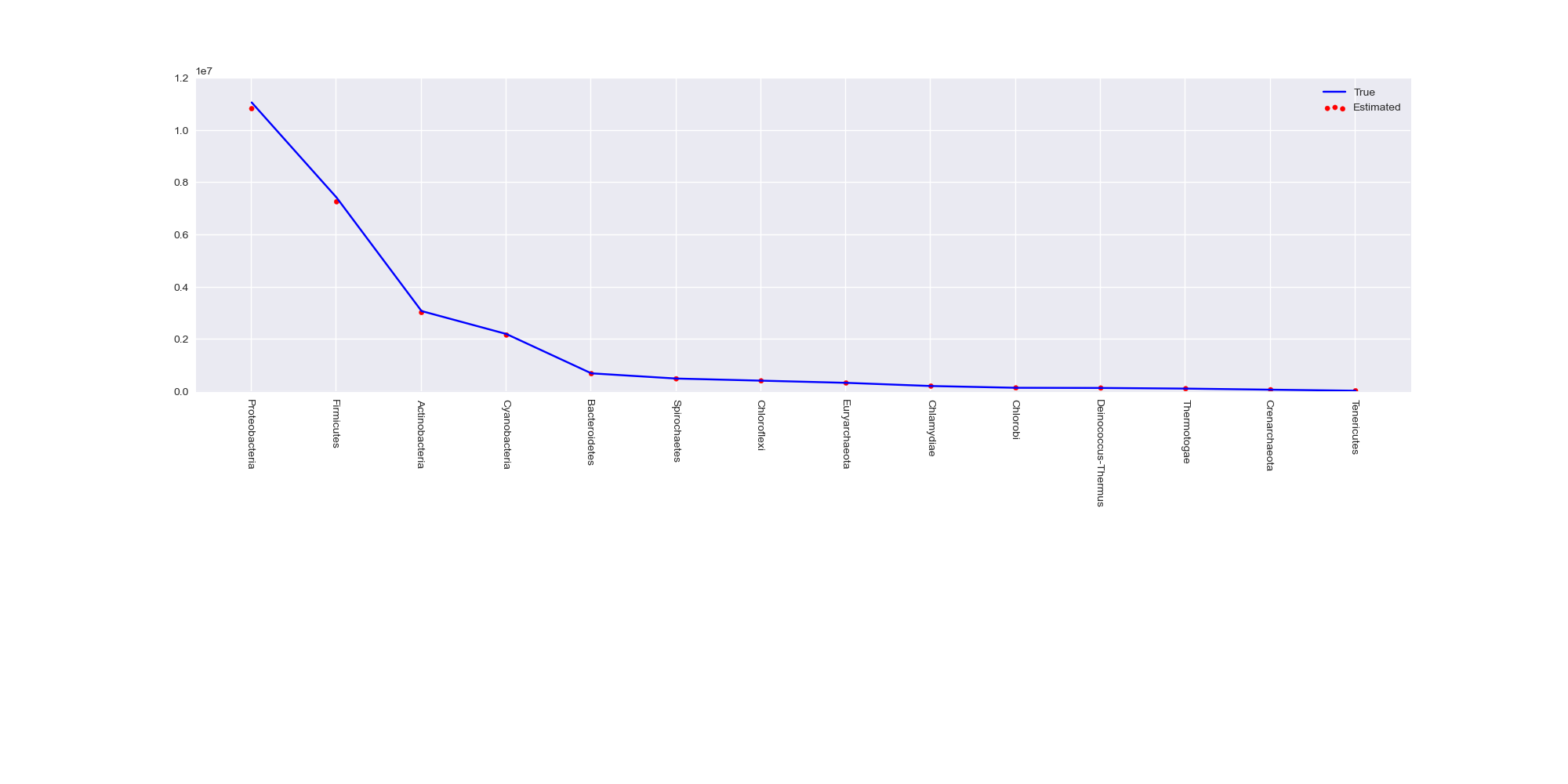}
       \label{fig:kallisto_phylum}
\end{subfigure}
\vskip -0.4in
    \begin{subfigure}[b]{\textwidth}
\includegraphics[width=\textwidth]{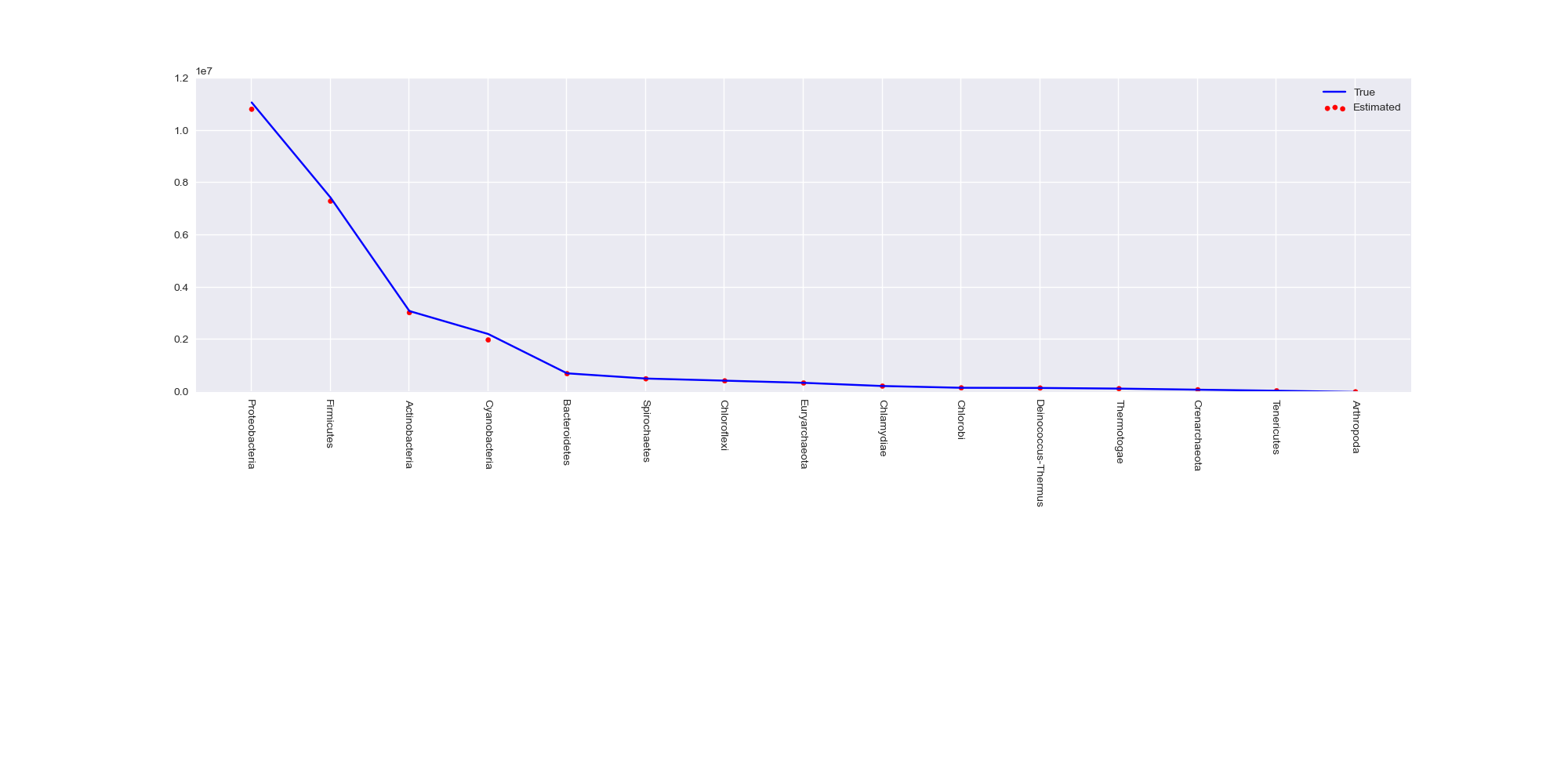}
       \label{fig:kraken_phylum}
\end{subfigure}
\vskip -0.4in
\begin{subfigure}[b]{\textwidth}
\includegraphics[width=\textwidth]{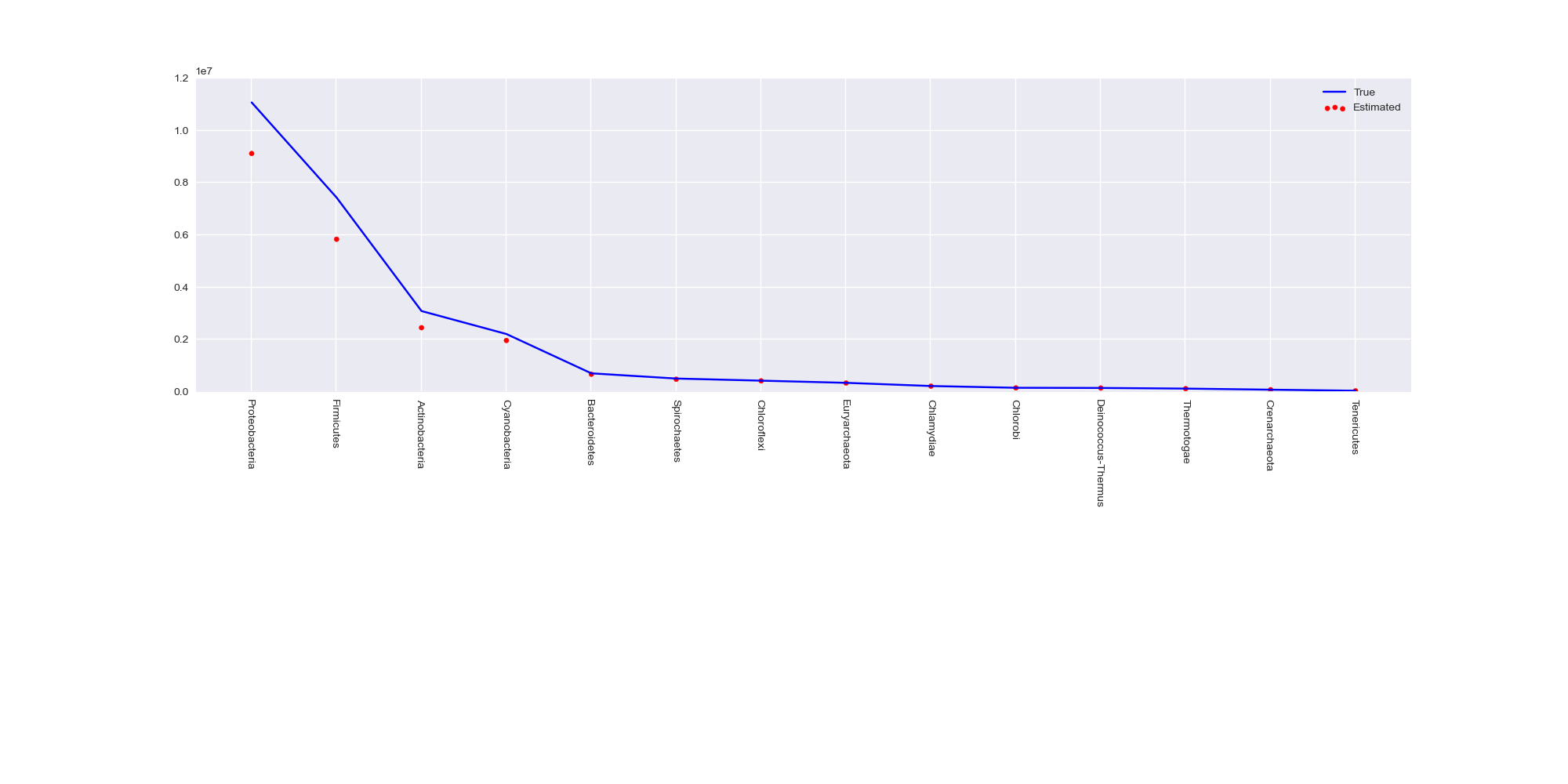}
       \label{fig:clark_phylum}
\end{subfigure}
\vskip -1.2in
\caption{Results of kallisto (top), Kraken (middle) and CLARK (bottom) on simulated reads from the i100+Martin dataset at the phylum level.}
\end{figure}

\end{document}